\documentclass[11pt,twoside]{article}

\usepackage{asp2006}
\usepackage{epsf}
\usepackage{psfig}
\usepackage{lscape}

\begin{document}
\title{Nearby Galaxies with Spitzer}   
\author{Daniela Calzetti$^1$, Robert C. Kennicutt$^2$, Daniel A. Dale$^3$, Bruce T. Draine$^4$, Charles W. Engelbracht$^5$, Armando Gil de Paz$^6$, Karl D. Gordon$^7$,  Juan C. Mu\~noz--Mateos$^6$, Eric J. Murphy$^8$ John D.T. Smith$^9$}   
\affil{$^1$~Department of Astronomy, University of Massachusetts -- Amherst; $^2$~Institute of Astronomy, University of Cambridge; $^3$Department of Physics and Astronomy, University of Wyoming; $^4$~Department of Astronomy, Princeton University; $^5$~Steward Observatory, University of Arizona; $^6$ Departamento de Astrof\''sica y CC. de la Atm\'osfera, Universidad Complutense de Madrid; $^7$~Space Telescope Science Institute;  $^8$ Spitzer Science Center, California Institute of Technology;  $^9$~Department of Physics and Astronomy, University of Toledo}    

\begin{abstract} 
We review the main advances brought by the Spitzer Space Telescope in the field of nearby galaxies studies, concentrating 
on a few subject areas, including: (1) the physics of the Polycyclic Aromatic Hydrocarbons that generate the mid--infrared features between $\sim$3.5~$\mu$m and $\sim$20~$\mu$m; (2) the use of the mid-- and far--infrared emission from galaxies as star formation rate indicators; and (3) the improvement of mid--infrared diagnostics to discriminate between thermal (star--formation)and non--thermal (AGN) emission in galaxies and galaxy centers. 
\end{abstract}


\section{Introduction}   
The main paradigm shift brought by the Spitzer Space Telescope in the investigation of nearby galaxies 
stems from a combination of high (for infrared observations) angular resolution ($\sim$2$^{\prime\prime}$--6$^{\prime\prime}$ 
at 8--24~$\mu$m), high sensitivity, and unprecedented mid--IR spectral--mapping capability. This has enabled 
the study of galaxies at the sub--kpc scale, thus resolving large morphological structures as traced by the dust 
emission and probing the ionization conditions of the interstellar medium (ISM) in those structures. 

Landmark advances brought by Spitzer observations include: a deeper understanding of the nature of the 
Polycyclic Aromatic Hydrocarbon (PAH) emission features (formerly  known as Unidentified Infrared Bands or Aromatic 
Features in Emission), and 
of the dependence of those features on the physical characteristics of the exciting stellar field (e.g., hardness of radiation) 
and the chemical characteristics of the ISM (e.g., metal content); a closer association between each dust 
emission component as a function of wavelength and the underlying heating stellar population, which, in turn, has led to 
better--calibrated mid-- and far--IR star formation rate (SFR) indicators; a better understanding of the physics underlying the infrared--radio correlation of galaxies \citep{Murphy2006,Murphy2008}; and important progress in the mid--infrared diagnostics to characterize and discriminate between star formation and active (non--thermal) galaxies and centers of galaxies. Many of these advances have enabled refinements of dust physics models, to better describe the shape and features of the dust infrared spectral energy distribution \citep[SED, e.g.,][]{Draine2007a}.

Why are we interested in the investigation of the infrared emission from nearby galaxies? Emission in the infrared fine structure lines  represents the main cooling channel for the ISM. Dust continuum emission represents about half of the global energy budget of the luminous component of the Universe \citep[e.g.][]{Hauser2001,Dole2006}. This latter statement is, furthermore, SFR-- and wavelength--dependent. As the SFR of galaxies increases, so does their dust opacity \citep[e.g.,][]{Wang1996, 
Calzetti2001, Sullivan2001}, and about 80\% of the UV light in the Universe is reprocessed by dust into the infrared. Clearly, 
a careful quantification of the impact of dust opacity on galaxy populations as a function of galactic parameters (morphology, 
luminosity, SFR, stellar population mix, gas content, metallicity, etc.) is a necessary step for recovering the stellar 
SEDs in those populations \citep[e.g.][]{Hopkins2006}. 

The Legacy from Spitzer in the field of nearby galaxies is, actually, still `work in progress' in just about all the aspects of dust and ISM emission mentioned above. Furthermore, the Warm Spitzer Mission 
is obtaining extensive observations of nearby galaxies at 3.6~$\mu$m and 4.5~$\mu$m, thus probing the low--mass stellar 
populations in a dust--extinction--free wavelength range; these data promise to yield major progress in our understanding of the formation and evolution of galactic structures (rings, bars, arms, etc.) and the extended disks. 

We present below a few highlights of the Spitzer results in the field of nearby galaxies; because of space limitations, 
however, this review will be mostly incomplete. 

\section{The Behavior of PAHs in Galaxies}

By building on initial ISO results, Spitzer observations have placed on a secure footing the strong dependence of the 
intensity of the mid--IR  PAH emission in galaxies on metallicity. The 8~$\mu$m--to--TIR luminosity ratio decreases by about an order of magnitude for a factor $\sim$10 decrease in metallicity, with a transition point 12$+$log(O/H)$\approx$8.1 \citep[e.g.,][]{Boselli2004,Madden2006,Engelbracht2005, Hogg2005,Galliano2005,Rosenberg2006,Draine2007b, Engelbracht2008,Marble2010}. This  decreasing trend is {\em on top} of the decrease in overall dust luminosity (at all wavelengths) that arises from the decreasing metal content. 

Most works compare different galaxies over a range of metallicity values, but a few also analyze radial profiles of  galaxies, finding basically the same result \citep{Gordon2008,Munoz2009}. \citet{Munoz2009} also suggests that the trend may revert for metallicity above 12$+$log(O/H)$\sim$9.0, although this result is tentative. 

The nature of the correlation between PAH strength and metallicity is still debated. Some authors \citep[e.g.][]{Galliano2008,Dwek2009} suggest it may be due to the delayed formation of the PAH, which are thought to form in the envelopes of carbon--rich AGB stars. In a young system, the first dust will emerge from supernovae (timescale$<$10~Myr), while AGB--produced dust will emerge at a later stage (timescale$\approx$1~Gyr). This scenario may be difficult to reconcile with the fact that low--metallicity systems in the local Universe contain stellar populations that are typically older than 2~Gyr \citep[see references in][]{Tosi2009}. However, low--mass systems are also thought to lose most of their metals during the sporadic events of star formation that characterize their typical star formation history \citep[e.g.,][]{Romano2006}. The latter may play a role in restoring a `delayed' 
dust enrichment of low--metallicity environments. The tentative reverse trend observed by \citet{Munoz2009} in metal--rich environments may also be accommodated within the evolutionary scenario for the PAH--metallicity dependence, a consequence of variations in the carbon relative abundance in AGB stars as a function of their mass and metallicity \citep{Galliano2008}. 

An alternate scenario to the `production' one calls for processing of PAHs by the radiation fields within which they are immersed \citep{Boulanger1988,Madden2006,Gordon2008,Engelbracht2008}. The original suggestion, based on the analysis of Milky Way nebulae, that the PAH emission is inversely correlated with the intensity of the stellar radiation field \citep{Boulanger1988} is not confirmed, at least on large scales, by the analysis of a range of galaxy environments \citep{Calzetti2007}. Processing of PAHs by hard radiation fields has, thus, been proposed as a mechanism for the observed correlation between PAH luminosity and metallicity 
\citep{Madden2006,Wu2006,Bendo2006,Smith2007,Gordon2008,Engelbracht2008}, since low--metallicity environments are generally characterized by harder radiation fields than high--metallicity ones \citep[e.g.][]{Hunt2010}. This suggestion agrees with the observation that PAHs are present in the PDRs surrounding HII regions, but are absent (likely destroyed) within the HII regions \citep{Helou2004,Bendo2006,Rellano2009}. A recent study of the SMC finds a high fraction of PAHs within molecular clouds \citep{Sandstrom2010}, thus complicating the interpretation of the formation/destruction mechanisms for these molecules. Overall, both production and processing may be driving the observed  trend \citep{Wu2006,Engelbracht2008,Marble2010}.

Outside of HII regions and other harsh environments, PAHs tend to be ubiquitous. In particular, they can be heated by the 
radiation from the mix of stellar populations that contribute to the general interstellar radiation field  \citep{Haas2002,Boselli2004,Peeters2004,Mattiola2005,Calzetti2007,Draine2007a,Bendo2008}. This suggests that PAH emission can be associated with evolved stellar populations unrelated to the current star formation in a galaxy.

\section{Monochromatic SFR Indicators in the Infrared}

The sensitivity and angular resolution of Spitzer have led to a new push in the derivation of SFR indicators based on a 
single IR band measurement, both for whole galaxies and for sub--galactic regions. The Spitzer  deep surveys of distant 
galaxies have provided a strong reason for calibrating such indicators, since measurements of distant galaxies 
in the IR are often limited to one or a few wavelength measurements. 

The rest--frame mid--infrared emission from dust in galaxies, in particular the emission detected 
in the 8~$\mu$m and 24~$\mu$m Spitzer bands (or the analogous ISO bands at 7~$\mu$m and 15~$\mu$m), has 
been analyzed by a number of authors  
\citep[][]{Roussel2001,Forster2004,Boselli2004,Calzetti2005, Wu2005, AlonsoHerrero2006, AlonsoHerrero2006b, PerezGonzalez2006,  Calzetti2007, Rellano2007, Zhu2008, Rieke2009, Kennicutt2009, Rellano2009,Salim2009}, and a general 
correlation (but also a number of caveats) between mid--IR infrared emission and SFR has been found.  

The restframe 8~$\mu$m dust emission generally shows a linear or almost--linear correlation with other SFR indicators 
\citep[extinction--corrected hydrogen emission lines, bolometric infrared emission, etc.;][]{Roussel2001,Forster2004,Calzetti2005,Wu2005,AlonsoHerrero2006b, Calzetti2007, Zhu2008,Kennicutt2009, Salim2009}. However, the calibration of the dust emission at 8~$\mu$m as a SFR 
indicator has been and still is a source of debate. The strong dependence on metallicity and/or hardness of the stellar 
radiation field, and the presence of PAH emission from dust heated by evolved stellar populations make such calibration 
difficult. The second contribution (the heating by evolved populations) is still unquantified, and we 
do not have a clear understanding of dependencies on morphology, stellar population mix, star formation rate, star 
formation intensity, etc. However, we can obtain a rough idea by converting the dust 8~$\mu$m mean  luminosity 
density, $\sim$1.2$\times$10$^7$~L$_{\odot}$~Mpc$^{-3}$, within the local 10~Mpc \citep[including emission from 
both PAHs and dust continuum, see][]{Marble2010} to a volume SFR density, using metallicity--dependent linear 
calibrations to SFR(8~$\mu$m$_{dust}$) from the data of \citet{Calzetti2007}. The result, $\rho_{SFR}$(8~$\mu$m$_{dust}$)$\sim$0.019~M$_{\odot}$~yr$^{-1}$~Mpc$^{-3}$, is roughly 30\%--60\% higher than the commonly accepted values for the 
SFR density in the local Volume \citep[see references in][]{Hopkins2006}; this indicates that on a global scale the 8~$\mu$m dust emission from galaxies traces the SFR with some excess, probably due to heating by evolved stellar populations \citep[see, also,][]{AlonsoHerrero2006b}. On a galaxy--by-galaxy basis or, worse, on a sub--galactic region--by--region basis, such additional heating could be even more important. 

The restframe 24~$\mu$m emission is more closely associated than the 8~$\mu$m emission with the dust heated by young, massive stars \citep{Helou2004,Calzetti2005,Rellano2009}, and should, therefore, provide a more accurate SFR indicator. The 
relation with SFR is, however, non--linear \citep{AlonsoHerrero2006, PerezGonzalez2006, Calzetti2007, Rellano2007,Kennicutt2009}, 
which indicates, for galaxies up to a luminosity L(TIR)$\sim$5$\times$10$^{10}$~L$_{\odot}$, that increasing SFR produces 
both larger dust emission and higher dust temperatures. The scatter in the data is sufficiently large that linear fits 
through the data can be drawn, yielding fairly consistent calibrations among different authors: SFR(24)$\sim$2.0$\times$10$^{-43}$~L(24), with a dispersion in the calibration constant of about 40\% for a \citet{Kroupa2001} stellar IMF \citep{Wu2005,Zhu2008,Rieke2009,Calzetti2010}, with the SFR in M$_{\odot}$~yr$^{-1}$ and L(24) in erg~s$^{-1}$. For galaxies with luminosity higher than L(TIR)$\sim$5$\times$10$^{10}$~L$_{\odot}$, a non--linear correction to this relation is necessary, owing to dust self--absorption in these systems \citep{Rieke2009}. 

Infrared--based calibrators, however, become inaccurate tracers of SFR in low metallicity/dust systems. A more effective way to trace SFRs across the full range of galaxy properties is to combine a tracer of obscured SFR (e.g., infrared) with a tracer of unobscured SFR \citep[e.g., UV or optical,][]{Calzetti2007,Kennicutt2007,Bigiel2008,Zhu2008,Kennicutt2009,Calzetti2010}. For galaxies:
\begin{eqnarray}
SFR&=& 5.45\times 10^{-42} [L(H\alpha)_{obs} + 0.020 L(24)],\ \ \ \ \ \ \ \ \ \ \ \ \ \ \ \ \ \ L(24)< 4\times 10^{42},\nonumber \\
         &=&5.45\times 10^{-42} [L(H\alpha)_{obs} + 0.031 L(24)], \ \ \ \ \ \ \ \ \ \ \ \ \ \ \ \ \ \  4\times10^{42}\le L(24)< 5\times 10^{43},\nonumber \\
          &=& 1.70\times 10^{-43} L(24) \times [2.03\times 10^{-44} L(24)]^{0.048}\ \ \ \ \ \ \ \ \ \ L(24)\ge 5\times10^{43},
\end{eqnarray}
with SFRs and luminosities in units of M$_{\odot}$~yr$^{-1}$ and erg~s$^{-1}$, respectively [L(24)=$\nu$L($\nu$)$_{24~\mu m}$; L(70), below, is defined similarly]. 

Dust emission at wavelengths longer than 24~$\mu$m also shows a generally linear correlation with SFR above $\sim$0.1--0.3~M$_{\odot}$~yr$^{-1}$, but with a dispersion about the mean that increases with wavelength \citep{Calzetti2010}. The dispersion is about 25\% (factor $\sim$2) larger at 70~$\mu$m (160~$\mu$m) than at 24~$\mu$m. Independent analyses of the 70~$\mu$m emission from star--forming regions within galaxies, however, suggest that the emission at this wavelength is better (more tightly) correlated  with SFR than either 8~$\mu$m, 24~$\mu$m or 160~$\mu$m emission  \citep{Lawton2010,Li2010}. The discrepancy may be due to the presence, in the integrated light of galaxies, of contribution to the infrared emission from dust heated by diffuse, evolved stellar populations; this contribution may be around 30\%--40\% at 70~$\mu$m \citep{Li2010} and larger at longer wavelengths \citep{Draine2007b,Calzetti2010}. A better understanding of the  contribution of the evolved stellar population to the dust heating in galaxies awaits the higher angular resolution data of the Herschel Space Telescope. For galaxies, the calibration at 70~$\mu$m is proposed as \citep{Calzetti2010}: SFR(70~$\mu$m)$\sim$5.9$\times$10$^{-44}$~L(70). 

\section{The Physics of the ISM}

Mid-infrared spectroscopy by Spitzer has revolutionized our understanding of the physical characteristics of the interstellar medium within nearby galaxies. The literature is sufficiently vast to merit a separate review, and only a few salient points will be mentioned here.

Spitzer observations of silicate emission and absorption in Type~1 and 2 Seyferts, respectively, have provided strong support for the unification model for Active Galactic Nucleus (AGN) galaxies \citep{Sturm2005, Siebenmorgen2005, Hao2005}.  Mid-infrared spectra of ULIRGs and QSOs have helped constrain the AGN fractional contribution to their bolometric luminosities \citep{Veilleux2009,Goulding2009}.  Spectroscopy from Spitzer has led to the improvement of mid-infrared diagnostics that characterize and distinguish between ULIRGs, AGN, and star--forming galaxies \citep{Sturm2006,Dale2006,OHalloran2006, Brandl2006,Spoon2007,Armus2007,Dudik2007, Hunter2007, Farrah2007, Dale2009}.  

For instance, the ratio between the PAH features at 7.7~$\mu$m and 11.3~$\mu$m decreases for increasing hardness of the radiation field in AGNs, in contrast to the near--universal ratio observed in star--formation--dominated galaxy regions \citep{Smith2007}. In a similar vein, the line ratios [FeII](25.99~$\mu$m)/[NeII](12.81~$\mu$m) and [SiII](34.82~$\mu$m/[SIII](33.48~$\mu$m)  are generally higher in AGNs than in star--formation--dominated regions, an effect that  may be due to variations in the depletion factors of Si and Fe onto dust grains, or the ionization characteristics of the nebular gas around AGNs, or to X--ray photoionization processes \citep{Dale2009}. 


Spitzer mid--IR spectroscopy has been instrumental in establishing that the low PAH abundance in low--metallicity galaxies is {\em not} due to the molecules being more highly ionized and/or dehydrogenated than in higher metallicity galaxies \citep{Smith2007}, although there is tentative evidence that they could be characterized by smaller sizes \citep{Hunt2010}. 
H$_2$ lines have been used to trace shocks and the excitation temperatures and masses of warm molecular hydrogen in galaxies \citep{Devost2004,Higdon2006,Appleton2006,Roussel2007,Hunter2007,Johnstone2007,Brunner2008}. Finally, the 
abundances of neon and sulfur derived from mid--infrared spectroscopy have been compared with those from optical lines to establish the impact of dust obscuration within nearby galaxies \citep{Wu2008, BernardSalas2009}.

Finally, Spitzer has established that many elliptical galaxies show unusual mid-infrared spectra \citep{Kaneda2005, Kaneda2008,Bregman2008}  that may be related to the presence of X-ray emission from low luminosity AGN.  \citet{Smith2007} suggest that the quiescent environments within ellipticals offer favorable conditions in which to observe these unusual spectra, since the mid-infrared spectra are not dominated by the effects of star formation typically seen in spiral and irregular galaxies.  

\acknowledgements 
This work is based in part on observations made with the Spitzer Space Telescope, which is operated by the Jet Propulsion Laboratory, California Institute of Technology under a contract with NASA.



\begin{thebibliography}{}
\bibitem[Alonso--Herrero et al.(2006a)]{AlonsoHerrero2006} Alonso--Herrero, A., Rieke, G.H., Rieke, M.J., Colina, L., Perez-Gonzalez, P.G., \& Ryder, S.D. 2006a, \apj, 650, 835
\bibitem[Alonso--Herrero et al.(2006b)]{AlonsoHerrero2006b} Alonso--Herrero, A., Colina, L., Packam, C., Diaz--Santos, T., Rieke, G.H., Radomski, J.T., \& Telesco, C.M. 2006b, \apj, 652, L83
\bibitem[Appleton et al.(2006)]{Appleton2006} Appleton, P.N., Xu, K.C., Reach, W., Dopita, M.A., Gao, Y., Lu, N., Popescu, C.C., et al. 2006, \apj, 639, L51
\bibitem[Armus et al.(2007)]{Armus2007} Armus, L., Charmandaris, V., Bernard-Salas, J., Spoon, H.W.W., Marshall, J.A., Higdon, S.J.U. et al. 2007, \apj, 656, 148
\bibitem[Bendo et al.(2006)]{Bendo2006} Bendo, G.J., Dale, D.A., Draine, B.T., Engelbracht, C.W., 
Kennicutt, R.C., Calzetti, D., Gordon, K.D., Helou, G., Hollenbach, D., Li, AIgen, Murphy, E.J., et al. 
2006, \apj, 652, 283
\bibitem[Bendo et al.(2008)]{Bendo2008} Bendo, G.J., Draine, B.T., Engelbracht, C.W., Helou, G.,  Thornley, M.D., Bot, C., Buckalew, B.A., Calzetti, D., Dale, D.A., Hollenbach, D.J., et al. 2008 \mnras, 389, 629
\bibitem[Bernard--Salas et al.(2009)]{BernardSalas2009} Bernard-Salas, J., Spoon, H. W. W., Charmandaris, V., Lebouteiller, V., Farrah, D., Devost, D., et al. 2009, \apjs, 184, 230
\bibitem[Bigiel et al.(2008)]{Bigiel2008} Bigiel, F., Leroy, A., Walter, F., Brinks, E., de Blok, W.J.G.,  Madore, B., Thornley, M.D. 2008, \aj, 136, 2846
\bibitem[Boselli, Lequeux \& Gavazzi(2004)]{Boselli2004} Boselli, A., Lequeux, J.,
  \& Gavazzi, G. 2004, \aap, 428, 409
\bibitem[Boulanger et al.(1988)]{Boulanger1988} Boulanger, F., Beichmann, C., Desert, F.--X., Helou, G., Perault, M., \& Ryter, C. 1988, \apj, 332, 328
\bibitem[Brandl et al.(2006)]{Brandl2006} Brandl, B.R., Bernard-Salas, J., Spoon, H.W.W., Devost, D., Sloan, G.C., Guilles, S., 
et al. 2006,  \apj, 653, 1129
\bibitem[Bregman et al.(2008)]{Bregman2008} Bregman, J.D., Bregman, J.N., \& Temi, P. 2008,`The Second Annual Spitzer Science Center Conference: Infrared Diagnostics of Galaxy Evolution' (ASP Conf. Ser.), eds. R.--R. Chary, H.I. Teplitz, \& K. Sheth (San Francisco, CA: ASP), Vol. 381, 34
\bibitem[Brunner et al.(2008)]{Brunner2008} Brunner, G.,Sheth, K., Armus, L., Wolfire, M., Vogel, S., Schinnerer, E., Helou, G., et al. 2008, \apj, 675, 316
\bibitem[Calzetti(2001)]{Calzetti2001} Calzetti, D. 2001, \pasp, 113, 1449
\bibitem[Calzetti et al.(2005)]{Calzetti2005} Calzetti, D., Kennicutt, R.C.,
Bianchi, L., Thilker, D.A., Dale, D.A., Engelbracht, C.W., Leitherer, C., 
Meyer, M.J., Sosey, M.L., Mutchler, M.  et al. 2005, \apj, 633, 871
\bibitem[Calzetti et al.(2007)]{Calzetti2007} Calzetti, D., Kennicutt, R.C., 
Engelbracht, C.W., Leitherer, C., Draine, B.T., Kewley, L., Moustakas, J., Sosey, M., Dale, D.A., 
Gordon, K.D.,  et al. 2007, \apj, 666, 870
\bibitem[Calzetti et al.(2010)]{Calzetti2010} Calzetti, D., Wu, S.-Y., Hong, S., Kennicutt, R.C., Lee, J.C., Dale, D.A., 
Engelbracht, C.W., van Zee, L., Draine, B.T., Hao, C.-N., et al. 2010, \apj, 714, 1256
\bibitem[Dale et al.(2006)]{Dale2006} Dale, D.A., Smith, J.D.T., Armus, L., Luckalew, D.A., Helou, G., Kennicutt, R.C., 
Moustakas, J. Roussel, H., Sheth, K., Bendo, G.J., et al. 2006, \apj, 646, 161
\bibitem[Dale et al.(2009)]{Dale2009} Dale, D.A., Smith,J.D.T., Schlawin, E.A., Armus, L., Buckalew, B.A., Cohen, S.A., 
Helou, G., Jarrett, T.H., Johnson, L.C., Moustakas, J., et al. 2009,  \apj, 693, 1821
\bibitem[Devost et al.(2004)]{Devost2004} Devost, D., Brandl, B.R., Armus, L., Barry, D.J., Sloan, G.C., Charmandaris, V.,
 Spoon, H.W.W., et al. 2004, \apjs, 154, 242
\bibitem[Dole et al.(2006)]{Dole2006} Dole, H., Lagache, G., Puget, J.-L., Caputi, K.I., Fernandez--Conde, N., Le Floc'h, E., 
Papovich, C., Perez--Gonzalez, P., Rieke, G.H., \& Baylock, M. 2006, \aap, 451, 417
\bibitem[Draine \& Li(2007)]{Draine2007a} Draine, B.T., \& Li, A. 2007, \apj, 657, 810. 
\bibitem[Draine et al.(2007)]{Draine2007b} Draine, B.T., Dale, D.A., Bendo, G., Gordon, K.D., Smith, 
J.D.T., Armus, L., Engelbracht, C.W., Helou, G., Kennicutt, R.C., Li, A., et al.  2007, \apj, 633, 866.
\bibitem[Dudik et al.(2007)]{Dudik2007} Dudik, R.P., Weingartner, J.C., Satyapal, S., Fischer, J., Dudley, C.C., \& O'Halloran, B. 2007, \apj, 664, 71 
\bibitem[Dwek et al.(2009)]{Dwek2009} Dwek, E., Galliano, F., \& Jones, A. 2009, `Cosmic DustÑNear and Far' (ASP Conf. Ser.), ed. T. Henning, E. GrŸn, \& J. Steinacker (San Francisco, CA: ASP), in press (arXiv:0903.0006)
\bibitem[Engelbracht et al.(2005)]{Engelbracht2005} Engelbracht, C.W., Gordon, K.D., Rieke, G.H., Werner, M.W., Dale, D.A., \& Latter, W.B.  2005, ApJ, 628, 29
\bibitem[Engelbracht et al.(2008)]{Engelbracht2008} Engelbracht, C.W., Rieke, G.H., Gordon, K.D., 
Smith, J.T.D., Werner, M.W., Moustakas, J., et al. 2008, \apj, 685, 678
\bibitem[Farrah et al.(2007)]{Farrah2007} Farrah, D., Bernard-Salas, J., Spoon, H.W.W., Soifer, B.T., Armus, L., Brandl, B.,  et 
al. 2007, \apj, 667, 149
\bibitem[F\"orster Schreiber et al.(2004)]{Forster2004} F\"orster Schreiber, N.M.,
  Roussel, H., Sauvage, M., \& Charmandaris, V., 2004, \aap, 419, 501
\bibitem[Galliano et al.(2005)]{Galliano2005} Galliano, F., Madden, S.C., Jones, A.P., Wilson, C.D., 
\& Bernard, J.-P. 2005, \aap, 434, 867  
\bibitem[Galliano et al.(2008)]{Galliano2008} Galliano, F., Dwek, E., \& Chanial, P. 2008, \apj, 672, 214
\bibitem[Gordon et al.(2008)]{Gordon2008} Gordon, K.D., Engelbracht, C.W., Rieke, G.H., Misselt, K.A., Smith, J.D.T., \& 
Kennicutt, R.C. 2008, \apj, 682, 336
\bibitem[Goulding \& Alexander(2009)]{Goulding2009} Goulding, A.D. \& Alexander, D.M. 2009, \mnras, 398, 1165
\bibitem[Haas, Klaas \& Bianchi(2002)]{Haas2002} Haas, M., Klaas, U., \&
Bianchi, S. 2002, \aap, 385, L23
\bibitem[Hao et al.(2005)]{Hao2005} Hao, L., Spoon, H.W.W., Sloan, G.C., Marshall, J.A., Armus, L., Tielens, A.G.G.M., et al. 2005,  \apj, 625, L75
\bibitem[Hauser \& Dwek(2001)]{Hauser2001} Hauser, M.G., \& Dwek, E. 2001, ARAA, ....
\bibitem[Helou et al.(2004)]{Helou2004} Helou, G., Roussel, H., Appleton, P.,
Frayer, D., Stolovy, S., Storrie--Lombardi, L., Hurst, R., Lowrance, P., et
al. 2004, \apjs, 154, 253
\bibitem[Higdon et al.(2006)]{Higdon2006} Higdon, S.J.U., Armus, L., Higdon, J.L., Soifer, B.T., \& Spoon, H.W.W. 2006, \apj, 648, 323
\bibitem[Hogg et al.(2005)]{Hogg2005} Hogg, D.W., Tremonti, C.A., Blanton, M.R., Finkbeiner, D.P., 
Padmanabhan, N., Quintero, A.D., Schlegel, D., \& Wherry, N.  2005, \apj, 624, 162
\bibitem[Hopkins \& Beacom(2006)]{Hopkins2006} Hopkins, A.M., \& Beacom, J.F. 2006 \apj, 651, 142
\bibitem[Hunt et al.(2010)]{Hunt2010} Hunt, L.K. , Thuan, T.X., Izotov, Y.I., \& Sauvage, M. 2010, \apj, 712, 164
\bibitem[Hunter \& Kaufman(2007)]{Hunter2007} Hunter, D.A. \& Kaufman, M. 2007, \aj, 134, 721
\bibitem[Johnstone et al.(2007)]{Johnstone2007} Johnstone, R.M., Hatch, N.A., Ferland, G.J., Fabian, A.C., 
Crawford, C.S., \& Wilman, R.J. 2007, \mnras, 328, 1246
\bibitem[Kaneda et al.(2005)]{Kaneda2005} Kaneda, H., Onaka, T., \& Sakon, I. 2005, \apj, 632, L83
\bibitem[Kaneda et al.(2008)]{Kaneda2008} Kaneda, H., Onaka, T., Sakon, I., Kitayama, T., Okada, Y., \& T. Suzuki, T. 2008, \apj, 684, 270
\bibitem[Kennicutt et al.(2007)]{Kennicutt2007} Kennicutt, R.C., Calzetti, D., Walter, F., Helou, G.,m Hollenbach, D., Armus, L., Bendo, G., Dale, D.A., Draine, B.T., Engelbracht, C.W., et al. 2007a, \apj, 671, 333
\bibitem[Kennicutt et al.(2009)]{Kennicutt2009} Kennicutt, R.C., Hao, C., Calzetti, D., et al. 2009, \apj, 
703, 1672 
\bibitem[Kroupa(2001)]{Kroupa2001} Kroupa, P. 2001, \mnras, 322, 231
\bibitem[Lawton et al.(2010)]{Lawton2010} Lawton, B., Gordon, K.D., Babler, B., Block, M., et al. 2010, \apj, in press (astroph/1005.0036)
\bibitem[Li et al.(2010)]{Li2010} Li, Y., Calzetti, D., Kennicutt, R.C., Hong, S., Engelbracht, C.W., Dale, D.A., Mopustakas, J., Gordon, K.D., Braine, B.T., et al. 2010, in prep. 
\bibitem[Madden et al.(2006)]{Madden2006}  Madden, S..... et al.  2006, \aap,  446, 877
\bibitem[Marble et al.(2010)]{Marble2010} Marble, A.R., Engelbracht, C.W., van Zee, L., Dale, D.A., Smith, J.T.D., Gordon, K.D., Wu, Y., Lee, J.C., Kennicutt, R.C., Skillman, E.D., Johnson, L.C., Block, M., Calzetti, D., COhen, S.A., Lee, H., \& Schuster, M.D. 2010, \apj, submitted
\bibitem[Mattioda et al. (2005)]{Mattiola2005} Mattioda, A.L., Allamandola, L.J., \& Hudgins, D.M.  2005, \apj, 629, 1183
\bibitem[Mu\~noz--Mateos et al.(2009)]{Munoz2009} Mu\~noz--Mateos, J.C., Gil de Paz, A., Boissier, S., Zamorano, J., 
Dale, D.A., P\'erez--Gonzalez, P.G., Gallego, J., Madore, B.F., Bendo, G., Thornley, M.D., Draine, B.T., Boselli, A., 
Buat, V., Calzetti, D., Moustakas, J., \& Kennicutt, R.C. 2009, \apj, 701, 1965
\bibitem[Murphy et al.(2006)]{Murphy2006} Murphy, E.J., Helou, G., Braun, R., Kenney, J.D.P., Armus, L., Calzetti, D., 
Draine, B.T., Kennicutt, R.C., Roussel, H., Walter, F., et al. 2006, \apj, 651, L111
\bibitem[Murphy et al.(2008)]{Murphy2008} Murphy, E.J., Helou, G., Kenney, J.D.P., Armus, L., \& Braun, R. 2008, \apj, 678, 828 
\bibitem[O'Halloran et al.(2006)]{OHalloran2006} O'Halloran, B., Satyapal, S., \& Dudik, R.P. 2006, \apj, 641, 795
\bibitem[Peeters, Spoon \& Tielens(2004)]{Peeters2004} Peeters, E., Spoon, H.W.W., \& Tielens, A.G.G.M. 2004, \apj, 613, 986
\bibitem[Perez--Gonzalez et al.(2006)]{PerezGonzalez2006} Perez--Gonzalez, P.G., Kennicutt, R.C., Gordon, K.D., Misselt, K.A., Gil de Paz, A., Engelbracht, C.W., Rieke, G.H., Bendo, G.J., Bianchi, L., Boissier, S., Calzetti, D., Dale, D.A., et al.  2006, \apj, 648, 987
 \bibitem[Rela\~no et al.(2007)]{Rellano2007} Rela\~no, M., Lisenfeld, U., Perez-Gonzalez, P.G., 
  Vilchez, J.M., \& Battaner, E. 2007, \apj, 667, L141
 \bibitem[Rela\~no \& Kennicutt(2009)]{Rellano2009} Rela\~no, M., \& Kennicutt, R.C. 2009, \apj, 699, 1125
\bibitem[Rieke et al.(2009)]{Rieke2009} Rieke, G.H., Alonso-Herrero, A., Weiner, B.J., Perez--Gonzalez, P.G., Blaylock, M., Donley, J.L, \& Marcillac, D. 2009, \apj, in press (astroph/0810.4150)
\bibitem[Romano et al.(2006)]{Romano2006} Romano, D., Tosi, M., \& Matteucci, F. 2006, \mnras, 365, 759
\bibitem[Rosenberg et al.(2006)]{Rosenberg2006} Rosenberg, J.L., Ashby, M.L.N., Salzer, J.J., \& Huang, J.-S. 2006, \apj, 636, 742
\bibitem[Roussel et al.(2001)]{Roussel2001} Roussel, H., Sauvage, M.,
Vigroux, L., \& Bosma, A. 2001, \aap, 372, 427
\bibitem[Roussel et al.(2007)]{Roussel2007} Roussel, H., Helou, G., Hollenbach, D.J., Draine, B.T., Smith, J.D., Armus, L., Schinnerer, E., et al. 2007, \apj, 669, 959
\bibitem[Salim et al.(2009)]{Salim2009} Salim, S., Dickinson, M., Rich, R. M., Charlot, S., Lee, J. C., Schiminovich, D., Perez-Gonzalez, P. G., Ashby, M. L. N., Papovich, C. Faber, S. M., et al. 2009, \apj, 700, 161
\bibitem[Sandstrom et al.(2010)]{Sandstrom2010} Sandstrom, K.M., Bolatto, A.D., Draine, B.T., Bot, C., \& Staminirovix, S. 2010, \apj, 715, 701
\bibitem[Siebenmorgen et al.(2005)]{Siebenmorgen2005} Siebenmorgen, R., Haas, M., Kr\"ugel, E., \& Schulz, B. 2005, \aap, 436, L5
\bibitem[Smith et al.(2007)]{Smith2007} Smith, J.D.T., Draine, B.T., Dale, D.A., Moustakas, J., Kennicutt, 
R.C., Helou, G., Armus, L., Roussel, H., Seth, K., Bendo, G.J., et al.  2007, \apj, 656, 770
\bibitem[Spoon et al.(2007)]{Spoon2007} Spoon, H.W.W., Marshall, J.A., Houck, J.R., Elitzur, M., Hao, L., Armus, L., Brandl, B.R., \& Charmandaris, V. 2007, \apj, 654, L49
\bibitem[Sturm et al.(2005)]{Sturm2005} Sturm, E., Schweitzer, M., Lutz, D., Contursi, A., Genzel, R., Lehnert, M.D., Tacconi, L.J., et al. 2005, \apj, 629, L21
\bibitem[Sturm et al.(2006)]{Sturm2006} Sturm, E., Rupke, D., Contursi, A., Kim, D.-C., Lutz, D., Netzer, H., et al. 2006, \apj, 653, 
L13
\bibitem[Sullivan et al.(2001)]{Sullivan2001} Sullivan, M., Mobasher, B., Chan, B., Cram, L., Ellis, R., 
Treyer, M., \& Hopkins, A. 2001, \apj, 558, 72
\bibitem[Tosi(2009)]{Tosi2009} Tosi, M. 2009, \aap, 500, 157
\bibitem[Veilleux, et al.(2009)]{Veilleux2009} Veilleux, S., Rupke, D.S.N., Kim, D.-C., Genzel, R., Sturm, E., Lutz, D., Contursi, 
A., et al. 2009, \apjs, 182, 628
\bibitem[Wang \& Heckman(1996)]{Wang1996} Wang, B., \& Heckman, T.M. 1996, \apj,
  457, 645
\bibitem[Wu et al.(2005)]{Wu2005}  Wu, H., Cao, C., Hao, C.-N., Liu, F.-S., 
Wang, J.-L., Xia, X.-Y., Deng, Z.-G., \& Young, C. K.-S. 2005, \apj, 632, L79 
\bibitem[Wu et al.(2006)]{Wu2006}  Wu, Y., Charmandaris, V., Hao, L., Brandl, B.R., 
Bernard-Salas, J., Spoon, H.W.W., \& Houck, J.R.  2006, \apj, 639, 157 
\bibitem[Wu et al.(2008)]{Wu2008} Wu, Y., Bernard-Salas, J., Charmandaris, V., Lebouteiller, V., Hao, L., Brandl, B.R., \& 
Houck, J.R. 2008, \apj, 673, 193
\bibitem[Zhu et al.(2008)]{Zhu2008} Zhu, Y.-N., Wu, H., Cao, C., \& Li, H.-N. 2008, \apj, 686, 155
\end{thebibliography}
\end{document}